\documentclass[Crown,sageh,times]{sagej}

\usepackage{moreverb,url}

\usepackage[colorlinks,allcolors=red]{hyperref}

\newcommand\BibTeX{{\rmfamily B\kern-.05em \textsc{i\kern-.025em b}\kern-.08em
T\kern-.1667em\lower.7ex\hbox{E}\kern-.125emX}}

\def\volumeyear{2018}

\newcommand{\w}{{\mathbf{w}}}

\begin{document}

\runninghead{Urban Analysis of Philadelphia}

\title{Urban Vibrancy and Safety in Philadelphia}

\author{Colman Humphrey\affilnum{1} and Shane T. Jensen\affilnum{1} and Dylan S. Small\affilnum{1} and Rachel Thurston\affilnum{2}}

\affiliation{\affilnum{1}Department of Statistics, The Wharton School, University of Pennsylvania\\
\affilnum{2}Stantec}

\corrauth{Shane T. Jensen, Department of Statistics
The Wharton School,
University of Pennsylvania,
463 Huntsman Hall
3730 Walnut Street,
Philadelphia, PA, USA 19102}

\email{stjensen@wharton.upenn.edu}

\begin{abstract}
Statistical analyses of urban environments have been recently improved through publicly available high resolution data and mapping technologies that have been adopted across industries.  These technologies allow us to create metrics to empirically investigate urban design principles of the past half-century.  Philadelphia is an interesting case study for this work, with its rapid urban development and population increase in the last decade.  We outline a data analysis pipeline for exploring the association between safety and local neighborhood features such as population, economic health and the built environment.  As a particular example of our analysis pipeline, we focus on quantitative measures of the built environment that serve as proxies for {\it vibrancy}: the amount of human activity in a local area.  Historically, vibrancy has been very challenging to measure empirically.   Measures based on land use zoning are not an adequate description of local vibrancy and so we construct a database and set of measures of business activity in each neighborhood.   We employ several matching analyses to explore the relationship between neighborhood vibrancy and safety, such as comparing high crime versus low crime locations within the same neighborhood.   We find that neighborhoods with more vacancy are associated with higher crime but within neighborhoods, crimes tend not to be located near vacant properties.   We also find that longer term residential ownership in a local area is associated with lower levels of crime.  In addition, we find that more crimes tend to occur near business locations but businesses that are active (open) for longer periods are associated with fewer crimes.   As additional sources of urban data become available, our analysis pipeline can serve as the template for further investigations into the relationships between safety, economic factors and the built environment at the local neighborhood level.  
\end{abstract}


\maketitle

\section{Introduction} \label{section-introduction}

Throughout history there have been many perspectives on the approach to planning of cities, with a notable clash between dense, organically-formed urban spaces versus large-scale clearing and planning of ``superblocks" and automobile-centric layouts. The former perspective viewed city development as a social enterprise created by many hands, whereas the top-down central planning approach involved less input from the residents affected by city changes.  The urban renewal movement of the 1960s and 1970s is the largest example of this latter effort, but the same mentality still drives many current development decisions.   

One historical motivation for top-down urban renewal projects was the idea that cities were over-crowded.  \cite{Win65} discusses both positive and negative perspectives on the effects of population density in urban settings.  \cite{Sim71} argues that the emotional stress caused by high population density produces negative attitudes and hostility among the populace. \cite{VerTay80} find both positive and negative effects of population density and suggest that population size is a more important factor for attitudes and behavior in urban environments.  

Earlier responses to anti-density rhetoric and the challenges of urban living during the industrial age resulted in code regulations, restrictive land use zoning, and sometimes large scale clearing of entire neighborhoods.  During the age of urban renewal, dense urban environments were demolished and replaced by trending architectural works, civic monuments and tree lined boulevards intended to reduce population density and ease automobile traffic, along with large housing projects for displaced communities.   Over time, a large number of these projects failed to attract pedestrian activity and resulted in high crime housing areas.  

In her seminal work {\it The Death and Life of Great American Cities} (1961), Jane Jacobs challenged the proponents of urban renewal and outlined several alternative hypotheses for sustaining successful urban environments.  Many of her ideas were based on her own anecdotal observations of urban residents, but can now be investigated quantitatively using recently available urban data.  

Jacobs was a pioneering voice for the concept of urban {\it vibrancy}, a measure of positive activity or energy in a neighborhood that makes an urban place enjoyable to its residents despite the challenges of urban living.   An important term coined by Jane Jacobs was  ``eyes on the street" which summarized her viewpoint that safer and more vibrant neighborhoods were those that had many people engaging in activities (either commercial or residential) on the street level at different times of the day \citep{Jac61}.   

This concept of eyes on the street has been more recently expressed as the ``natural surveillance" component of the {\it Crime Prevention through Environmental Design} movement \citep{Deu16}.   These contemporary theories argue that the likelihood of criminal activity is strongly linked to the presence or absence of people on the street.  As \cite{Deu16} states: ``Criminals do not like to be seen or recognized, so they will choose situations where they can hide and easily escape."   Policies which promote vibrancy and activity could potentially benefit crime prevention.  

The recent explosion in high resolution data on cities offers an exciting opportunity for quantitative evaluation of contrasting urban development perspectives as well as current urban planning efforts.  The goal of this paper is to outline a pipeline for data collection and analysis of the relationships between neighborhood safety, economic and demographic conditions and the built environment within urban environments.

As an example, we target our analysis pipeline towards a specific task: using high resolution data to create quantitative measures of the built environment that can serve as proxies for the human {\it vibrancy} of a local area.  We then investigate the association between these vibrancy measures and safety in the city of Philadelphia.    We focus on vibrancy measures based on land use as well as business activity, which follows the  ``natural surveillance" idea that the presence of open businesses encourages safety through the store front presence of both staff and customers.   

Past investigations into the built environment and safety includes \cite{RonMai91} which investigates the influence of liquor establishments on crime in Cleveland.  \cite{WeiBusLum04} explore the trajectories of crime over a fourteen year period on specific street segments in Seattle, while \cite{GroMcC12} examines the association between crime and specific characteristics of parks in Philadelphia.  

\cite{mac15} provides a recent comprehensive review of past research into the association between the built environment and safety, where many quasi-experimental studies have shown that changes in housing, zoning and public transit can help to manage crime.  In Section~\ref{businessmatching}, we will try to emulate a quasi-experimental setting in our own analysis by comparing locations within census block groups, thereby matching locations in terms of their economic and demographic characteristics. 

The effects of natural surveillance on neighborhood vibrancy can be both subtle and complicated.  The presence of a commercial business can encourage vibrancy through the presence of many people, or can give a sense of vacancy and isolation to an area if it is closed during a particular time of the day.   In order to get an accurate picture of whether commercial businesses help to encourage safety, we will need to examine whether or not those businesses are open and active, as we outline in Section~\ref{businessvibrancy}. 

Jane Jacobs also wrote about the positive benefits of residents that feel invested long term in their communities.  The defensible space hypothesis of \cite{New76} argues that an area is safer when people feel a sense of ownership and responsibility in their local community.   We will investigate this hypothesis through the relationship between the long term ownership tenure of residents and neighborhood safety. 
Philadelphia is an interesting case study for this work as it is currently encountering many contemporary issues in urban revival, population growth and desirability.   Recent work has shown that urban city centers are growing relative to their suburban counterparts in many areas of the country \citep{CouHan15}.  \cite{EllHorRee16} finds an association between population movement of high-income and college-educated households and declining crime rates in central city neighborhoods.  

We first outline our data collection for the city of Philadelphia in Section~\ref{section-data} and then explore the associations between safety and several economic, population and land use measures in Section~\ref{section-eda}.  To get a more detailed picture of neighborhood vibrancy, we compile a database and several measures of business vibrancy in Section~\ref{businessvibrancy}.  

In Section~\ref{businessmatching}, we employ several matching analyses to evaluate the association between business vibrancy, land use, long term ownership and safety within local neighborhoods.   We find several interesting associations between crime and our measures of vibrancy, land use, and ownership tenure, though we are not implying causal effects with these findings.  We conclude with a brief discussion in Section~\ref{section-discussion}. 

In order to encourage replication of our analyses and adaptation to other research questions, we have made the code and public data that were used in our analyses available as a github repository at: \\ {\small {\tt https://github.com/ColmanHumphrey/urbananalytics}}  

\section{Urban Data in Philadelphia} \label{section-data}

Our analysis will be based on the geographical units defined by the US Census Bureau.   Philadelphia county is divided into 384 census tracts which are divided into 1,336 block groups which are divided into 18,872 blocks.    Population and economic data is provided by the US Census Bureau, crime data is provided by the Philadelphia Police Department and land use data is provided by the City of Philadelphia.

A general theme of our urban work is that results can vary depending on the resolution level of the data. Most of our analyses will be done at the block group level which allows for the best interoperability between our economic and built environment data, but we also perform several analyses at the block level.   

Shape files from the US Census Bureau delineate the boundaries and area of each census block and block group.   Shape files from the City of Philadelphia delineate the boundaries and area of each lot in Philadelphia.  For the vast majority of lots in Philadelphia, the lot is entirely contained within a single Census Bureau block.  

\begin{figure}[ht]
\centering
\includegraphics[width=5.5in]{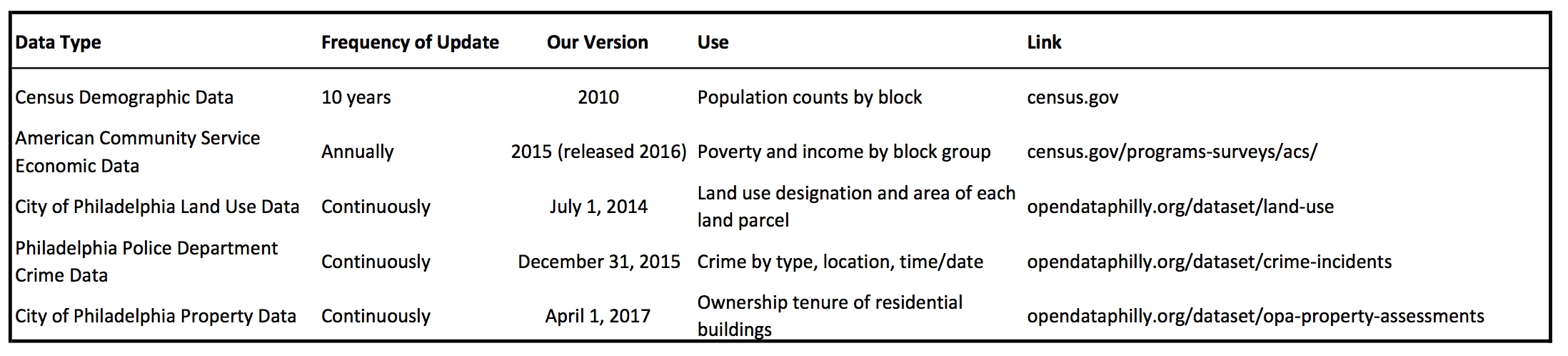}
\caption{Additional details about each data source used in our analysis.  For each data type that we used, we list the source of that data (with web links) and how we used that data.  We also list how frequently that data source is updated and when we specifically acquired that data source.}
\label{data-details-table}
\end{figure}

Figure~\ref{data-details-table} summarizes the data sources used in our analysis and we provide additional details for each data source in Sections~\ref{demographicdata}-\ref{crimedata}. 

\subsection{Population Data}\label{demographicdata}

Our population demographic data were pulled from the census website (\texttt{factfinder.census.gov}) by setting the geography as all 
blocks in Philadelphia and setting the data source as ``Hispanic or Latino Origin By Race'' (which is SF1 P5 in their database). The raw demographic data give the population count in each block from the 2010 census.    

Of the 18,872 blocks in Philadelphia, 4,558 have no residents (e.g. parks, industrial areas, etc.). At the block level, we restrict our analysis to blocks with at least 25 people, which gives 12,874 blocks that contain 98.9\% of the population.  At the block group level, we restrict our analysis to block groups with at least 400 people in them, which gives 1,325 block groups (out of 1,336) that contain 99.96\% of the population.  We calculate the population count and population density in each block group $i$ from the raw population data and using the area of each block group from the US Census Bureau shape files.

\subsection{Economic Data}\label{economicdata}

Our economic data were pulled from the American Community Survey on the census website (\texttt{factfinder.census.gov}): tables B19301 for income and C17002 for poverty, both from 2015.  These data are only available at the block group resolution level.
 
For each block group in Philadelphia, we have the per-capita income
and the fraction of the population in seven different brackets of
income-to-poverty-line ratios: $[0, 0.5), [0.5, 1), [1, 1.25), [1.25, 1.5), [1.5, 1.85), [1.85, 2),[2, \infty)$.
For interpretation, the $[0.5, 1)$ bracket represents families that have
income between $50\%$ of the poverty line and the poverty line.  

The poverty line threshold for each household is defined by the Census Bureau according to the size and composition
of the household.   As an example, a family of four with two children has a poverty line threshold of \$23,999.  We define a scalar poverty measure for each block group based on the weighted
sum of the proportion of block group households in each of the seven poverty brackets:
\[
\text{Poverty}_i = \sum_{q = 1}^7 w_q \, p_{i,q}
\]
where $p_{i,1}$ is the proportion of block group $i$ households in the lowest bracket $[0, 0.5)$ and $p_{i,7}$ is the proportion of block group $i$ households in the highest bracket $[2, \infty)$.   We employ linearly decreasing weights $\w = [1, 5/6, 4/6, 3/6, 2/6, 1/6, 0]$ to give highest weight to the brackets with highest poverty.   Our $\text{Poverty}_i$ metric varies from 0 to 1, with larger values implying higher poverty.   Maps by block group of population density, per-capita income and poverty in Philadelphia are given in Figure S1 of our supplementary materials.  

\subsection{Land Use Zoning Data}\label{landusedata}

Land use zoning data were downloaded from the City of Philadelphia.   The land use data consist of a shapefile that divides the city into approximately 560,000 lots and provides the area and registered land use zoning designation (commercial, residential, industrial, vacant, transportation, water, park, civic, recreation, culture, and cemetery) for each of these lots.   Almost all Philadelphia city lots are contained entirely within a single census block (and block group) which makes it easy to calculate the proportion of land in a census block (or block group) that is designated with a particular land use.  

Note that we combine the ``commercial business" and ``commercial consumer"
into a single {\it commercial} designation, and all three ``residential"
categories into a single {\it residential} designation.   For the rest of this paper, {\it mixed use} refers to the designation of ``commercial / residential mixed".  We create several land use measures from these zoning designations.  First, we calculate the fraction of area in each geographic unit (either block or block group) $i$ that is designated as `Vacant',   
\[
\text{Vacant.Prop}_{i} = \frac{Area_i (\text{Vacant})}{Area_i}
\]
Second, we calculate the ratio of the area in each geographic unit (either block or block group) $i$ that is commercial versus residential, 
\small
\[
\text{ComRes.Prop}_i = \frac{Area_i(\text{Commercial})}{Area_i(\text{Commercial}) + Area_i(\text{Residential})}
\]
\normalsize
Finally, we calculate the proportion of every block or block group that is designated as mixed use, 
\small
\[
\text{MixedUse.Prop}_i = \frac{Area_i(\text{Mixed Use})}
{Area_i}
\]
\normalsize
These land use zoning metrics provide our first set of proxy measures for the vibrancy of a local neighborhood.  Further details about the land use designations and our created land use metrics are given in Figure S2 of our supplementary materials.  

Philadelphia's zoning procedures were revised in 2012 (\url{http://www.phila.gov/li/Pages/Zoning.aspx}).   Our zoning data were downloaded in June 2014, and all of our analyses are based on that snapshot. While most of the city's zoning remains unchanged, lots can be rezoned through applications on a continual basis.

\subsection{Property Data}\label{propertydata}

Property data are made available by the City of Philadelphia in the {\it Property} dataset on \texttt{opendataphilly.org}.   These data contain the estimated market value, size, age and various other characteristics for each property in the city of Philadelphia.  

With these data, we focus our analysis pipeline on long term residential ownership, which we estimate by tabulating the time period since the last sale for every residential property in Philadelphia.  For any location in the city, we can calculate the {\it average ownership tenure}: the time period since the last sale for every residential property around that specified location.  

\subsection{Crime Data}\label{crimedata}

Crime data for Philadelphia come from the Philadelphia Police Department through the \texttt{opendataphilly.org} website.  The data contain the date and time of each crime as well as the location in terms of latitude and longitude (WGS84 decimal degrees). Each crime is categorized into one of several types: Homicide, Sexual, Robbery, Assault, Burglary, Theft, Motor Theft, Arson, Vandalism, or Disorderly Conduct.  

For our subsequent analysis, we combine these types into two super-categories of crimes: a. {\bf violent crimes} (Homicides, Sexual, Robbery and Assault) and b. {\bf non-violent crimes} (Burglary, Theft, Motor Theft, Arson, Vandalism, and Disorderly Conduct).   
Further details about the relative frequency and spatial distribution of crimes in Philadelphia are provided in Figure S3 of our supplementary materials.

\section{Exploring Neighborhood Factors Associated with Safety in Philadelphia} \label{section-eda}

\subsection{Association between Crime and Population}\label{popassociation}

We first examine whether population is associated with either violent or non-violent crimes in Philadelphia.   We find that population count is more strongly associated with both violent crime ($r = 0.26$) and non-violent crime ($r = 0.46$) than population density.  In fact, population density is not significantly associated with violent crime ($r = -0.01$), and negatively associated with non-violent crime ($r = -0.09$).  These correlations were calculated from a robust regression that downweights outlying values \citep{huber81}.  Plots of these relationships are provided in Figure S4 of our supplementary materials.  We also explored Poisson and Negative Binomial regressions but found that these alternative formulations did not give substantially different results.  

The lack of a strong positive association between population density and crime is notable in the context of historical hypotheses such as \cite{Sim71} which argue that high population density leads to negative attitudes and hostility.   In contrast, we find population size to be more strongly associated with crime, which supports the work of \cite{VerTay80} though this finding is specific to our focus on Philadelphia.   

To incorporate the association between crime and population count into our subsequent analyses, we define {\it excess violent crime} in each block group as the residuals from the robust regression of violent crime on population count.  Similarly, we define {\it excess non-violent crime} in each block group as the residuals from the robust regression of non-violent crime on population count.  We can interpret these excess crime (violent or non-violent) totals as the number of crimes in a block group beyond their expectation based on population count. 

\subsection{Association between Excess Crime and Economic Measures} \label{econassociation}

As outlined in Section~\ref{economicdata}, we focus on two measures of the economic health of each block group in Philadelphia: per-capita income and our constructed poverty metric.  We find a strong negative relationship between excess violent crime and income  ($r = - 0.44$) and a strong positive relationship between excess violent crime and poverty ($r = 0.59$).   We also find substantial non-linearity in the relationship between income and violent crime, with a strong linear relationship between violent crime and income for per-capita income below \$50,000 but very little relationship above per-capita income of \$50,000.  

These economic measures have a much weaker relationship with excess non-violent crime.  There is a weak negative association between per-capita income and excess non-violent crime ($r = -0.12$) and a weak positive association between poverty and excess non-violent crime ($r = 0.33$).   Plots of these relationships are provided in Figure S5 of our supplementary materials.

Together, these results suggest that per-capita income and poverty are strongly associated with excess violent crime but not excess non-violent crime, possibly because non-violent crimes are more likely to be crimes of opportunity occurring in areas located away from where the perpetrators reside.  Crimes of opportunity may be more driven by locations of businesses which helps to motivate our work in Sections~\ref{businessvibrancy}-\ref{businessmatching}. 

To incorporate the association between crime and these economic measures into our subsequent analysis, we now re-define {\it excess violent crime} in each block group as the residuals from the robust regression of violent crime on population count, per-capita income and poverty; Similarly for {\it excess non-violent crime}.   We can interpret these excess crime (violent or non-violent) totals as the number of crimes in a block group beyond their expectation based on population count, income and poverty.  

\subsection{Association between Excess Crime and Land Use Zoning} \label{landuseassociation}

Up to this point, we have focussed on the relationship between safety and features based on the population and economic health of each neighborhood.   However, our primary interest is the influence of the {\it built environment} of the neighborhood on safety, which could inform future urban planning initiatives.  One type of data that we have pertaining to the built environment are the land use zoning designations for each lot in the city of Philadelphia (Section~\ref{landusedata}), from which we created three measures of the ``vibrancy" in each block group: the fraction of vacant land, the fraction of mixed use land, and the ratio of commercial area to residential area. 

We find a moderately strong positive relationship between vacant proportion and excess violent crime ($r = 0.2$) and between vacant proportion and excess non-violent crime ($r = 0.2$).  We also find moderately strong positive relationships between mixed use proportion and excess violent crime ($r = 0.23$) and between mixed use proportion and non-violent crime ($r = 0.23$).  We find stronger positive relationships between commercial vs. residential proportion and excess violent crime ($r = 0.42$) and between commercial vs. residential proportion and excess non-violent crime ($r = 0.65$).  Plots of these relationships are provided in Figure S6 of our supplementary materials.

The association we find between vacant proportion and safety is relevant to recent studies of the effect of ``greening" vacant lots on neighborhood safety \citep{BraCheMac11}.  That study found that the ``greening" of vacant lots was associated with a reduction in gun assaults and vandalism.

The strong positive relationship we find between commercial proportion and crime is also interesting in the context of contemporary theories of urbanism.   As we describe in Section~\ref{section-introduction}, the ``eyes on the street" theory of \cite{Jac61} and ``natural surveillance" theory of \citep{Deu16} argue that safer neighborhoods are those that have greater presence of people on the street achieved through a mixing of commercial and residential properties.  Our preliminary findings do not support the idea that a mix of commercial and residential land use is associated with increased safety. 

However, we must concede that land use zoning designations are a low resolution indication of vibrancy that only indicate intended use.  In particular, the zoning designation of a lot as commercial does not provide insight into whether the commercial enterprise located there contributes positively or negatively to vibrancy in the area or whether that commercial enterprise is open or closed during times when crimes tend to be committed.   This missing information motivates our investigation of more detailed measures of neighborhood vibrancy based on business data in the following Section~\ref{businessvibrancy}.

\section{Urban Vibrancy Measures based on Business Data} \label{businessvibrancy}

As discussed in Section~\ref{landuseassociation}, measures based on land use zoning designations are an insufficient summary of the vibrancy of a neighborhood.   We can not evaluate whether a mix of commercial and residential properties promotes safety without first establishing what types of business enterprises are present in lots zoned for commercial use and when those businesses are active. To that end, we outline our manual assembly and curation of a database of Philadelphia businesses, as well as the construction of several measures of business vibrancy from those data.

\subsection{Business Data}\label{vibrancy-googledata}

We have manually assembled a database of Philadelphia businesses by scraping three different web sources: Google Places, Yelp, and Foursquare.  Each of these sources provide the GPS locations for a large number of businesses in Philadelphia, as well as opening hours for a subset of those businesses.    

The most difficult issues with assembling this business database are: 1. integrating these three data sources and removing overlapping businesses and 2. categorizing all businesses into a small set of {\it business types}.   Table~\ref{table-business-counts} gives the number of businesses and the number of those businesses where we have opening hour information.  We also give counts of the total number of businesses and the number of businesses with opening hours in the union of all three data sources (removing duplicates between data sources).  

\begin{table}[ht!]
\caption{Number of businesses and number with opening hours from each data source. }
\centering
    \begin{tabular}{ p{3.5cm} | p{1.5cm} p{1.5cm} p{1.5cm} | p{1.5cm} }
      & Google & Yelp & Foursquare & Union \\ \hline
      Total businesses & $34,768$ & $12,534$ & $40,331$ & $72,020$\\
      Businesses with hours & $12,346$ &  $7,728$ &  $7,022$ & $19,140$\\
    \end{tabular}
\label{table-business-counts}
\end{table}

Each data source has its own categorization for businesses, with Google using about a hundred categories and Yelp and Foursquare each using closer to a thousand categories.   Out of this myriad of business categorizations, we created ten {\it business types}: Cafe ($4,166$), Convenience ($1,481$), Gym ($1,273$), Institution ($24,489$), Liquor ($316$), Lodging ($461$), Nightlife ($5,108$), Pharmacy ($799$), Restaurant ($7,909$),
and Retail ($31,147$). The values in parentheses are the total number of businesses in each business type. 

A particular business can belong to multiple business types, e.g. a restaurant that also sells liquor to go.  Most of these business types are self-explanatory, but we need to clarify a few details.   The cafe type includes cafes, bakeries and coffee shops that are not full restaurants.   The restaurant type also includes meal delivery and meal take out businesses.  Institution is a broad type that includes banks, post offices, churches, museums, schools, police and fire departments, as well as many others.

\subsection{Measures of Business Vibrancy} \label{busvibmeasures}

We use our assembled business data to create several high resolution measures of business vibrancy at any particular location in the city of Philadelphia.   We want these measures to encapsulate whether a given location has a concentration of a businesses of a particular type, and whether those businesses are active storefronts (i.e. open) during times of the week when crimes tend to be highest.  We focus on two high crime windows that have a disproportionately large number of crimes (both violent and non-violent) relative to other times of the week.  These two high crime windows are {\it weekday evenings}: 6-12pm Monday-Friday and {\it weekend nights}: 12-4am Saturday-Sunday.   

The first set of measures of business vibrancy we consider are simply the total number of businesses of each {\it business type} near to any particular location in the city, as we expect that some business types will be more associated with safety than others.  We also 
want to take into account whether those businesses are active storefronts in the sense of being open.  In particular, we are interested in whether a given location has businesses of a particular type (e.g. cafes) that are open longer than expected.   

We first establish a {\it consensus} number of open hours for each business type by calculating the average open hours across all businesses of that type in Philadelphia.   For each business in Philadelphia where we have open hours, we can then calculate its {\it excess} number of open hours relative to the consensus for its business type.    Building upon these calculations, the second set of measures of business vibrancy we consider are the average excess hours of businesses near to any particular location in the city.    

Thus, for any particular location in Philadelphia, we have two sets of business vibrancy measures: the number of businesses of each business type and the average excess hours of each business type.  The excess hour measures can be calculated over the entire week or just within the high crime windows mentioned above.   In Section~\ref{businessmatching}, we evaluate the association between these business vibrancy measures and both excess violent and non-violent crimes within the local neighborhoods of Philadelphia.  

\section{Association between Business Vibrancy, Land Use, Ownership and Safety} \label{businessmatching}

Our goal is evaluating the association between our created business vibrancy measures and safety at the local neighborhood level, while controlling for the characteristics of those neighborhoods.   We will control for neighborhood characteristics by focusing our analyses on comparing pairs of locations within each block or within each block group.  Underlying this strategy is an assumption that the census blocks or block groups are small enough that different locations within these geographic units should be highly similar with regards to the demographic and economic measures we examined in Sections~\ref{popassociation} and \ref{econassociation}.  

In Section~\ref{high.vs.low} below, we focus on pairs of locations within blocks where one location has the highest frequency of crimes and the other location has the lowest frequency of crimes within that block.   We then examine these within-block pairs to see if there are differences our business vibrancy measures between the ``high crime" vs. ``low crime" locations.   As an additional study, we also compared crime totals between a location with a business that has longer hours versus another location in that same block group with a business that has shorter hours.  Details of this study are given in Section 3 of our supplementary materials.

\subsection{Comparing ``High Crime" vs. ``Low Crime" Locations}\label{high.vs.low}

We first calculate the location with the highest crime frequency and the location with the lowest crime frequency within each block.   We perform this analysis on the census block level (rather than the block group level) in order to give an even higher resolution view of the association between vibrancy and safety.   For each block, we calculate our two sets of business vibrancy measures, the number of businesses of each business type and the average excess hours of each business type, in a 50 meter radius around the high crime and low crime locations in those blocks. 

Many blocks do not contain any businesses of a particular type near either the highest or lowest crime locations, which excludes those blocks from any comparisons involving that particular business type.  We further restrict ourselves to blocks where the highest crime and lowest crime locations are at least 100 meters apart so that the 50 meter radii around these locations do not overlap.

For each business type, we calculate a matched pairs t-statistic for differences in the business vibrancy measures around the low crime location minus the business vibrancy measures around the high crime location within each block.    If business vibrancy helps to deter crime and promote safety, then these differences in business vibrancy should be positive.  

\begin{figure}[ht!]
\centering
\includegraphics[width=5.5in]{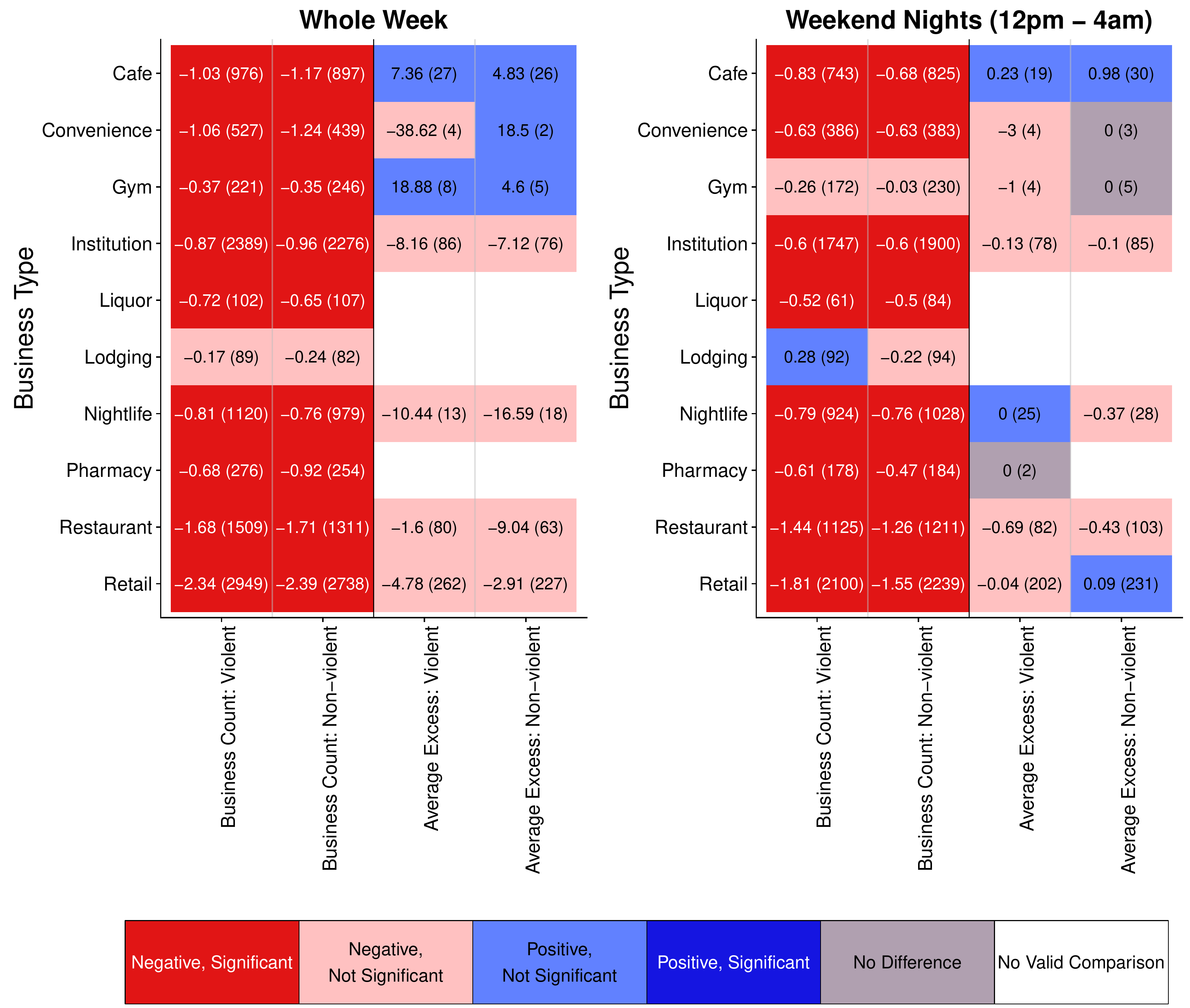}
\caption{Matched pair mean differences in measures of business vibrancy between high crime and low crime locations, calculated separately for each combination of crime type and business type.  Different panels are used to display the mean differences calculated over the entire week vs. just weekend nights. The significance threshold of $p = 0.05$ was Bonferroni-adjusted to account for multiple comparisons.  Values in parentheses are the number of blocks with valid comparisons for that business type.}
\label{fig.high.vs.low}
\end{figure}

In Figure~\ref{fig.high.vs.low}, we display the matched mean differences in the two business vibrancy measures (the number of businesses of each business type and the average excess hours of each business type) between the low crime and high crime within-block locations.  We calculate differences for locations based on violent crimes and locations based on non-violent crimes.  The significance threshold for these t-statistics was Bonferroni-adjusted to account for the number of comparisons being performed.  Figure~\ref{fig.high.vs.low} provides these comparisons for the entire week as well as the {\it weekend nights} time window.  Due to limited space, we do not display the comparisons for {\it weekday evenings}, but the results are highly similar to {\it weekend nights}.  

We see in Figure~\ref{fig.high.vs.low} that the difference in number of businesses is significantly negative (red) for both violent and non-violent crimes for essentially all business types, most strongly retail stores and restaurants.  This result suggests that there are significantly more businesses around the higher crime locations than the lower crime locations.  

However, we also observe in Figure~\ref{fig.high.vs.low} that for some of these business types, such as gyms and cafes, there are positive differences (blue) for our average excess hours metric, which implies that businesses are open longer around the low crime location compared to the high crime location.  These differences are not as significant, but we still see evidence of an interesting and subtle finding: more crimes tend to occur near cafes and gyms but fewer crimes tend to occur near cafes and gyms that are open longer. 

We can also compare our original land use zoning measures of vibrancy from Section~\ref{landuseassociation} in a 50 meter radius between these high and low crime locations.   We again calculate differences for locations based on violent crimes and locations based on non-violent crimes, but now the differences are based on our three land use vibrancy measures: the fraction of vacant land, the fraction of mixed use land and the ratio of commercial area to residential area.  We also calculate differences in average ownership tenure of residential properties (as outlined in Section~\ref{propertydata}) between the high crime and low crime locations.  

\begin{figure}[ht!]
\centering
\includegraphics[width=5.5in]{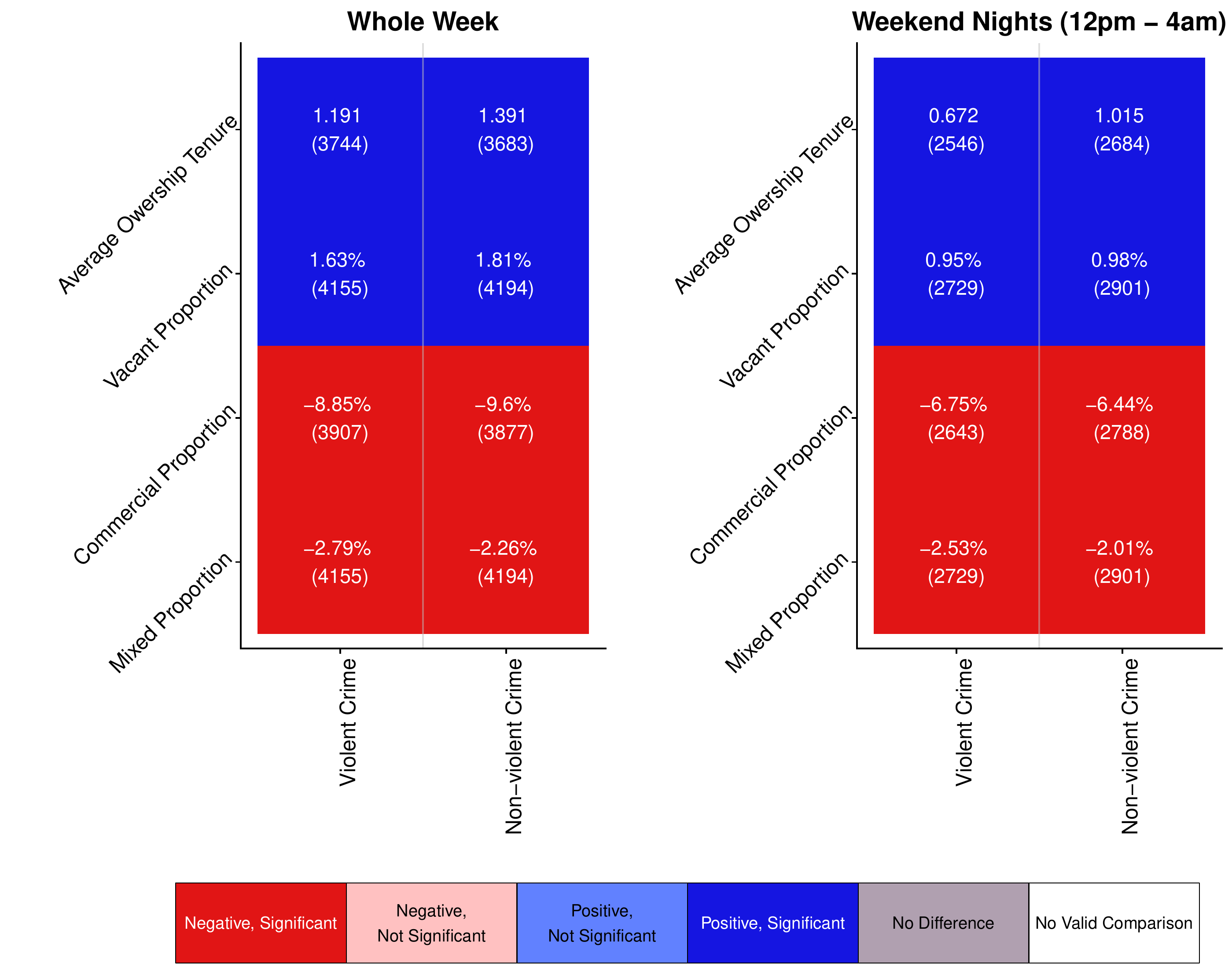}
\caption{Matched pair mean differences in average ownership tenure and three measures of land use zoning vibrancy between high crime and low crime locations.  Different panels are used to display the mean
differences calculated over the entire week vs. just weekend nights.  The significance threshold of $p = 0.05$ was Bonferroni-adjusted to account for multiple comparisons.  Values in parentheses are the number of block groups with valid comparisons for that business type.}
\label{bus.vib.landuse}
\end{figure}

Figure~\ref{bus.vib.landuse} gives the matched mean differences in average ownership tenure and our three land use vibrancy measures between the low crime and high crime within-block group locations.   We display these comparisons for the entire week as well as the {\it weekend nights} time window.  The comparisons for {\it weekday evenings} are not displayed due to limited space but the results are highly similar to {\it weekend nights}.  

In Figure~\ref{bus.vib.landuse}, we see very strong negative differences for mixed proportion and commercial vs. residential proportion, both of which strongly suggests that there is more mixed and commercial zoning near to the high crime locations. This association between commercial enterprises and crime was also observed in Section~\ref{landuseassociation} and motivated our development of more detailed measures of business vibrancy in Section~\ref{businessvibrancy}.

We also see very strong positive differences for the vacant land proportion which suggests the presence of more vacant land near to low crime locations compared to the high crime locations.  This finding is notable when compared to the positive association between vacant proportion and crime that we found in Section~\ref{landuseassociation}.    Together, those two findings suggest that neighborhoods with more vacant properties overall have higher crime but when looking within those neighborhoods, crimes tend not to be located near vacant properties.   These results are especially interesting given the mixed effects on crime from the ``greening" of vacant lots in the study by \cite{BraCheMac11}.  

We also see in Figure~\ref{bus.vib.landuse} that average ownership tenure of residential properties is significantly longer around the low crime location compared to the high crime location for both violent and non-violent crimes.  This finding provides some support for the defensible space hypothesis \cite{New76} that longer term residential investment in the community (which we estimate with average ownership tenure) is associated with lower crime.

\subsection{Summary of Business Vibrancy and Safety}

Our analysis pipeline for studying the association between land use, business vibrancy and safety in Section~\ref{high.vs.low} above (and Section 3 of our supplementary materials) has produced several findings that could inform current evaluations of contemporary theories in urban planning.  

First, we find that more crimes occur near business locations but that some types of businesses (such as cafes and gyms) that are open for longer periods are associated with fewer crimes.   Second, we find that although neighborhoods with more aggregate vacancy have higher crime (Section~\ref{landuseassociation}), when comparing locations within each neighborhood, crimes tend not to be located near vacant properties.  Third, we find significantly longer residential ownership around the low crime locations compared to the high crime locations. 

Another important observation from Figure \ref{fig.high.vs.low} is the substantial heterogeneity in the association between business vibrancy and crime both across different business types and different time windows.   The power of both studies was compromised by small sample sizes as there are only a limited number of block groups that permit a pair of comparable locations.     The associations between land use zoning and safety in Figure~\ref{bus.vib.landuse} are more significant due to much larger sample sizes of locations for these comparisons.    Clearly, the associations between safety and neighborhood vibrancy are subtle, heterogeneous, and in need of even higher resolution studies to fully understand.

\section{Discussion}\label{section-discussion}

The recent availability of high resolution data on cities provides a tremendous opportunity for sophisticated quantitative evaluation of historical and current urban development.    To aid these efforts, this paper outlines a framework for data collection and analysis of the associations between safety, economic and demographic conditions and the built environment within local neighborhoods.    We used this framework for a specific task: the creation of quantitative measures of ``vibrancy" based on the built environment of a neighborhood and exploration of the association between these vibrancy measures and neighborhood safety.  

We find that population density is not strongly associated with either violent or non-violent crime, which argues against the theory of \cite{Sim71}.  We find that population count is a more important predictor of crime, which supports the work of \cite{VerTay80}.     We also explored the association between crime and economic measures as well as measures of vibrancy derived from land use zoning data, but found that these measures were not an adequate summary of the local commercial vibrancy of an area.  

To address vibrancy at a higher resolution, we constructed several measures of business vibrancy and employed matching of locations within block groups to evaluate the relationship between business vibrancy and safety.  Our business vibrancy measures (number of businesses and average excess hours of businesses) are designed to be proxies for the  ``eyes on the street" concept of \cite{Jac61}.  

Our results suggest that more crimes tend to occur near business locations but that businesses of some types that are active (open) for longer periods are associated with fewer crimes. We also find that the overall proportion of vacancy in a neighborhood is associated with higher crime but that within a neighborhood, crimes tend to not occur as often near to vacant properties.  Finally, we find that longer term residential investment in the community (as measured by average ownership tenure of residential properties) is associated with lower levels of crime.   

However, these preliminary findings should not be interpreted as causal effects.  For example, underlying community effects could be driving both the longer business hours and and lower crime rates in our observed associations.  Rather, we consider our findings to be opportunities for further in-depth investigation.  More direct and high resolution measures of foot traffic or human activity would certainly improve measures of urban vibrancy.  For example, \cite{YueZhuYeh17} use mobile phone data to estimate vibrancy in Shenzhen, China.  However, these types of direct data sources are not currently publicly available for the city of Philadelphia.  Measures of human activity from geographically-linked social media usage are another promising research direction.  For example, \cite{LaiCheLan17} use geo-coded Twitter data to measure human activity around London Underground stations.  

As further studies deepen our understanding of the role of vibrancy as an indicator of safety, we can consider public policy initiatives that encourage vibrancy by promoting multiple use spaces as well as the potential deregulation of closing times and noise curfews in order to allow businesses to experiment with longer opening hours.      More generally, we encourage the adaptation of our analysis pipeline to other research questions within the urban analytics community.  The code and public data that were used in our analyses is available as a github repository at: {\small {\tt https://github.com/ColmanHumphrey/urbananalytics}}

\section{Acknowledgments}

This research was supported by a grant from the Wharton Social Impact Initiative (WSII) at the University of Pennsylvania.

\nocite{r_itself}
\nocite{rpkg_tidycensus}
\nocite{rpkg_dplyr}

\bibliographystyle{SageH}
\bibliography{references}
   
\newpage

\bigskip

\begin{center}
{\Large {\bf Supplementary Materials for ``Urban Vibrancy and Safety in Philadelphia"}}
\end{center}

\bigskip

\section{Maps of Urban Data in Philadelphia} \label{section-data}

\begin{figure}[ht]
\renewcommand\thefigure{S1}
\centering
\includegraphics[width=5.5in]{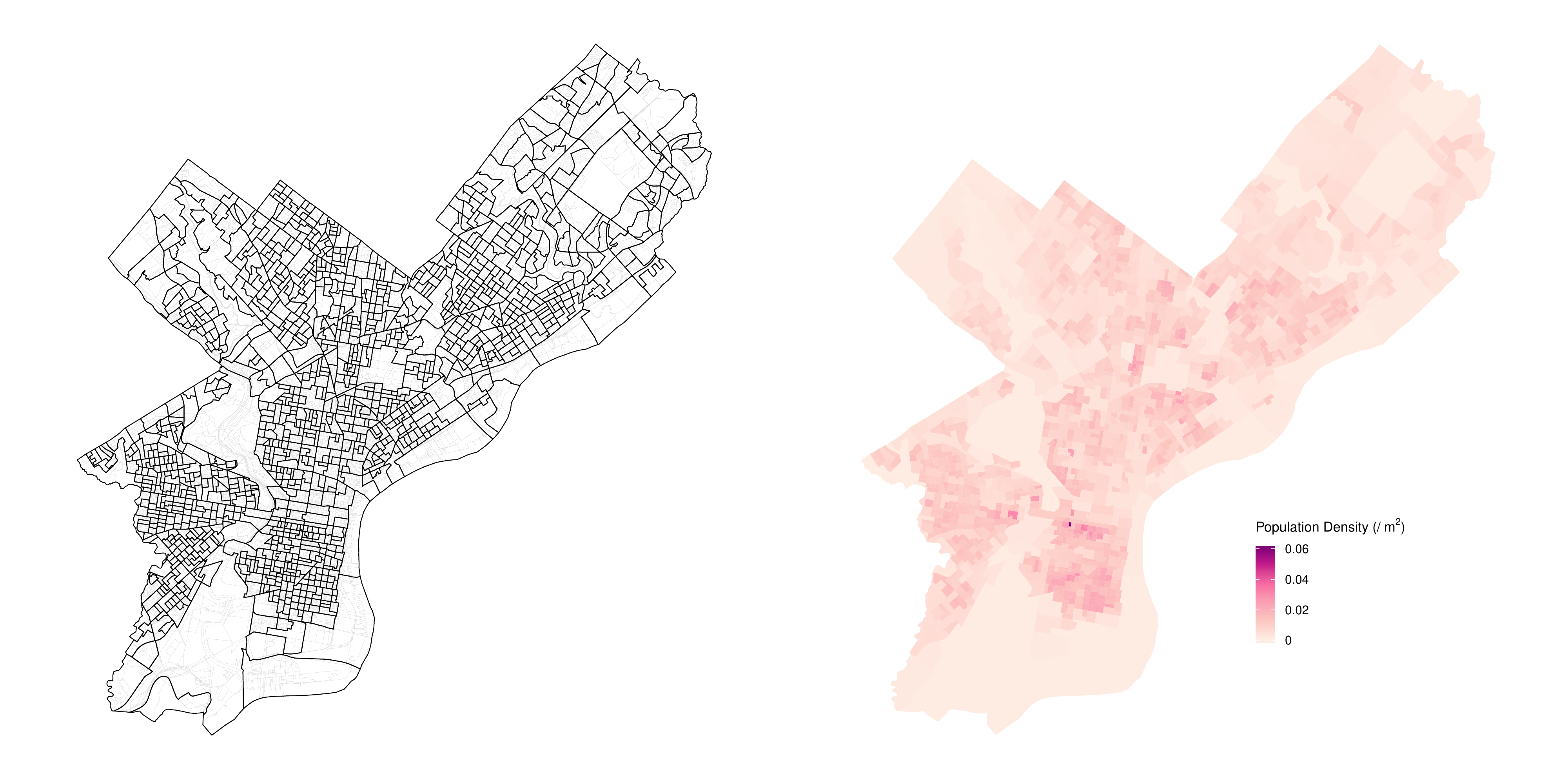}
\includegraphics[width=5.5in]{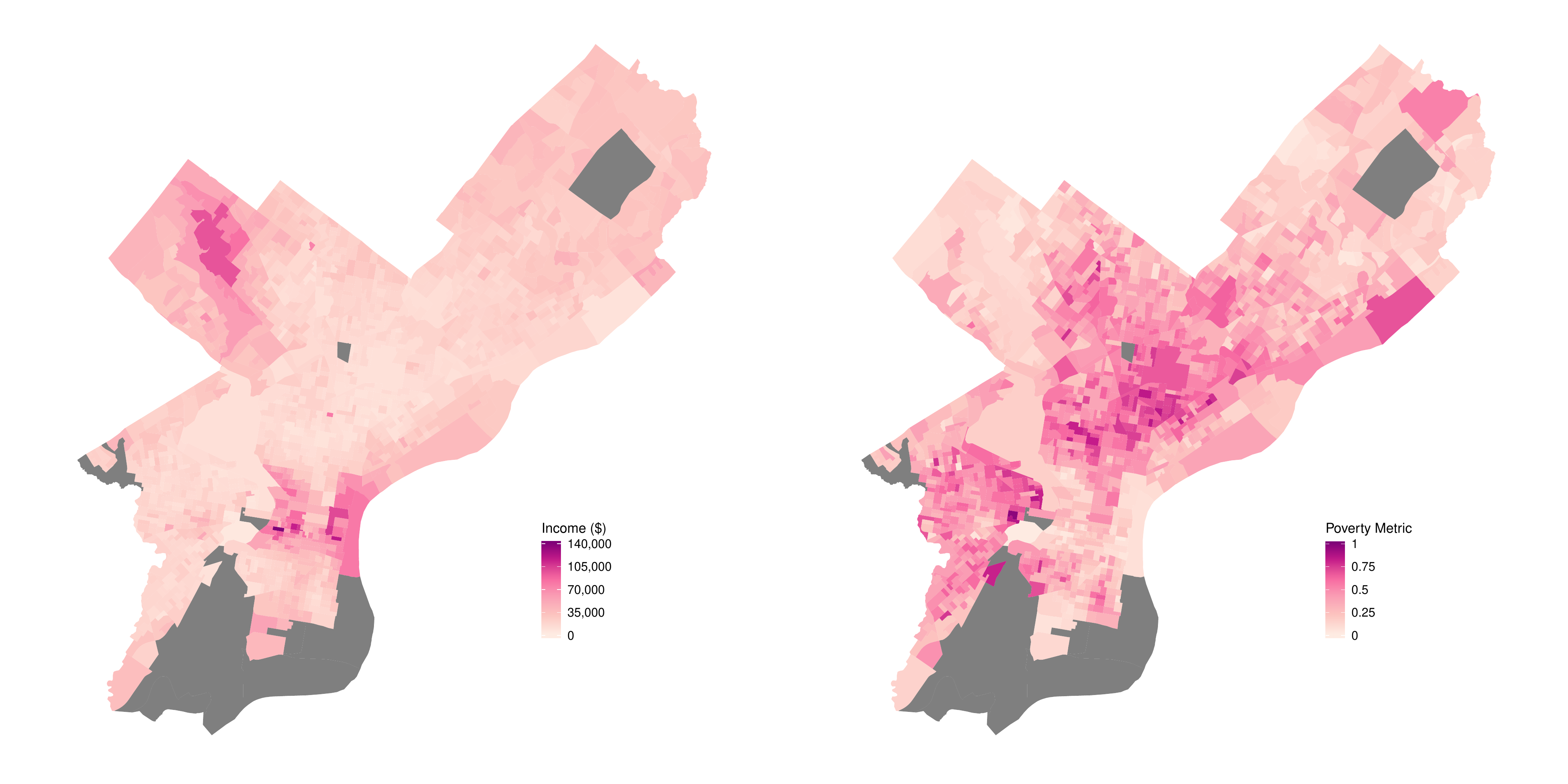}
\caption{{\bf Top Left:} Map of Philadelphia divided into block groups (black lines) by US Census Bureau.  {\bf Top Right:} Population density by block group in Philadelphia {\bf Bottom Left:} Per-capita income by block group in Philadelphia {\bf Bottom Right:} Poverty metric by block group in Philadelphia.  Block groups that are colored grey do not have enough residents for the US Census Bureau to report economic data for those block groups.}
\label{map-philadelphia-population-economic}
\end{figure}

Figure~\ref{map-philadelphia-population-economic} gives a map outlining the 1,336 block groups in Philadelphia and the population density, per capita income and poverty metric for each of those block groups.   Philadelphia county is divided into 384 census tracts which are divided into 1,336 block groups which are divided into 18,872 blocks.   The area of the Philadelphia census blocks has an average of 0.00756 square miles (median of 0.00372 square miles) with a standard deviation of 0.0235 square miles.  The area of the Philadelphia block groups has an average of 0.107 square miles (median of 0.0482 square miles) with a standard deviation of 0.323 square miles.  

\smallskip

\begin{figure}[ht]
\renewcommand\thefigure{S2}
\centering
\includegraphics[width=4.4in]{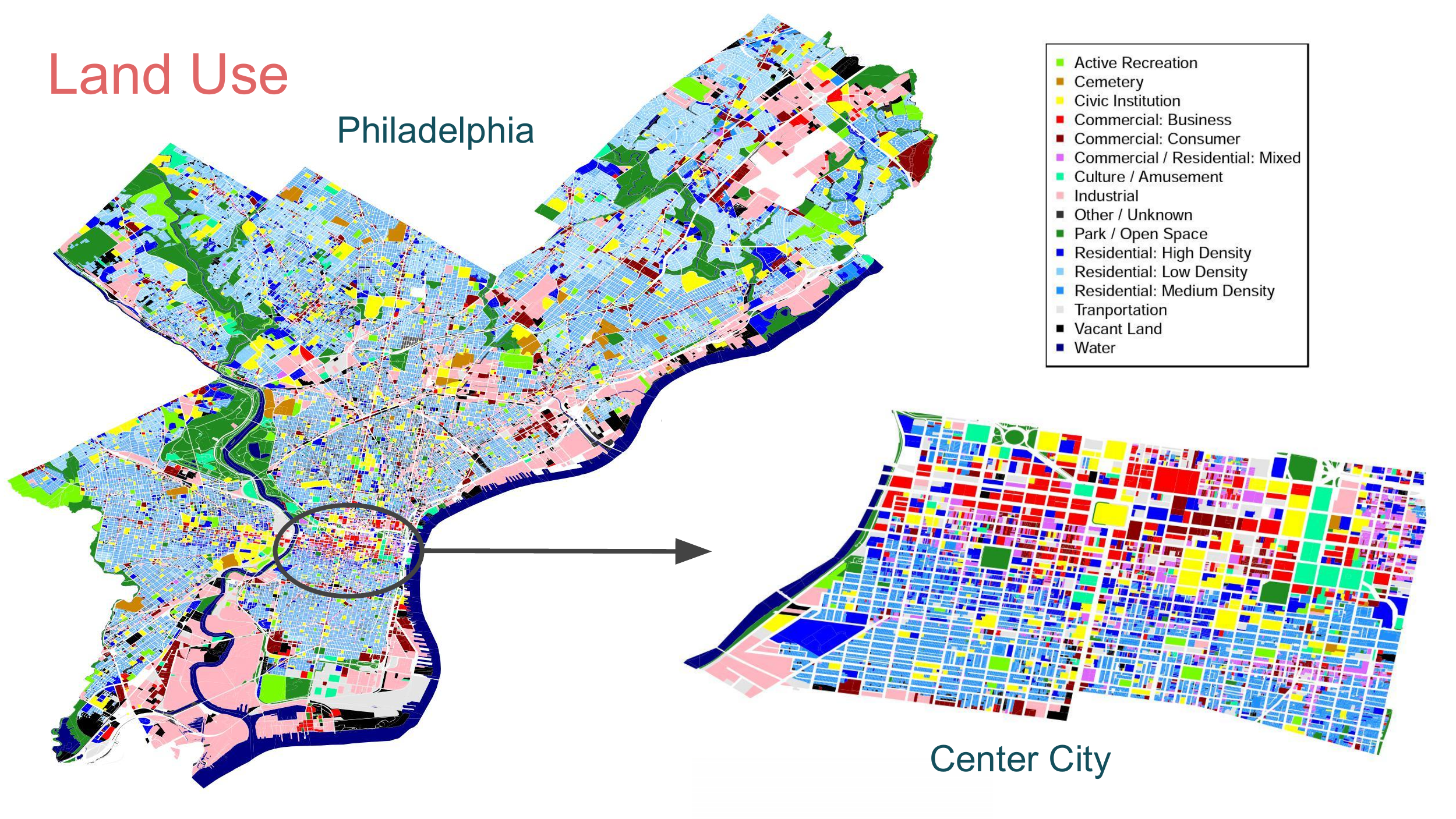}
\includegraphics[width=5in]{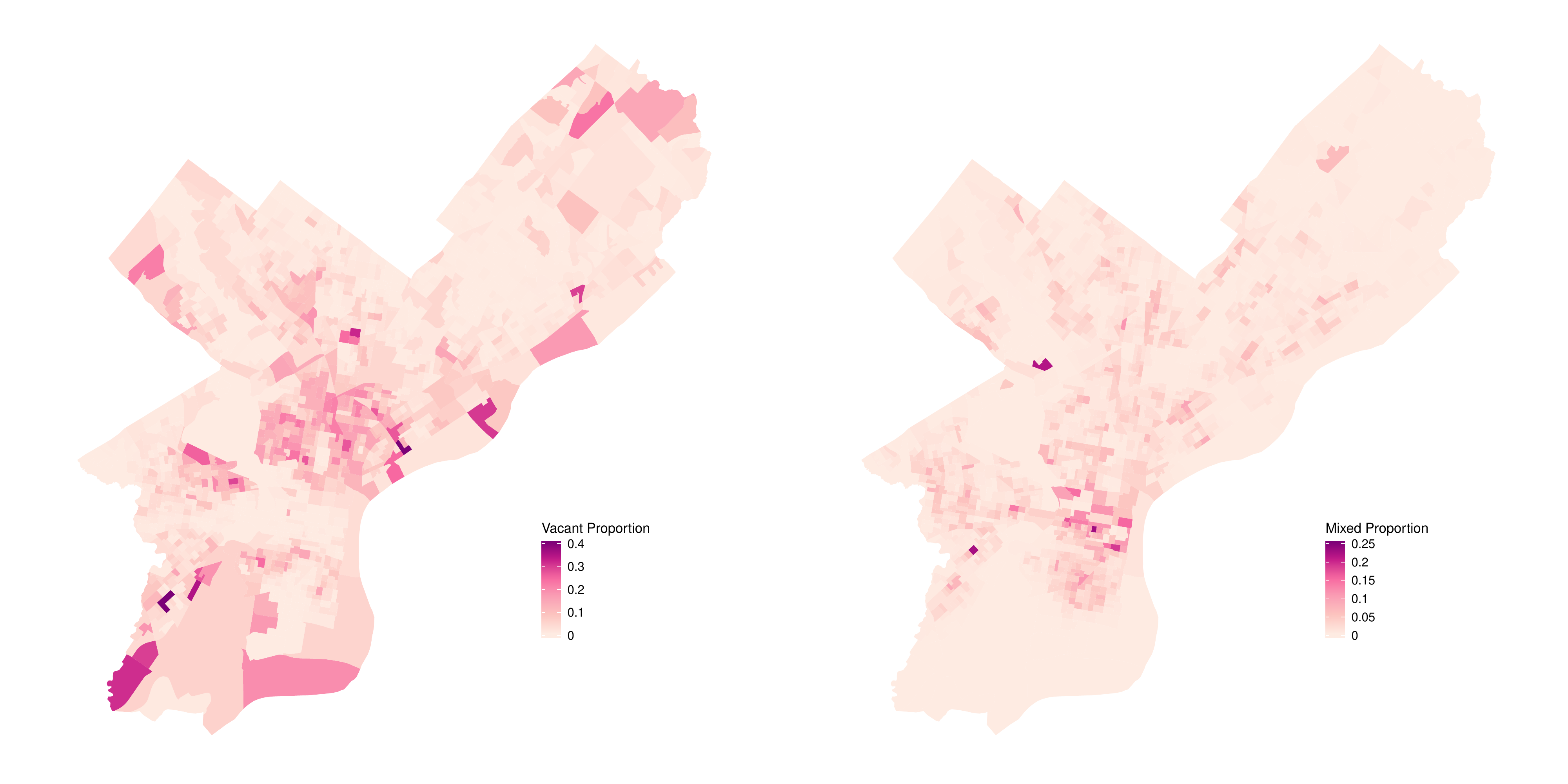}
\caption{{\bf Top:} Land use designations for overall Philadelphia and the center city neighborhood.  These designations were used to calculate the {\bf Bottom Left:} Vacant Proportion and {\bf Bottom Right:} Mixed Use Proportion for each block group. }
\label{map-landuse}
\end{figure}

\noindent
Figure~\ref{map-landuse} gives the land use designations for the city of Philadelphia with a focus in on the neighborhood of center city, as well as the distribution by block group of the vacant proportion (left) and mixed use proportion (right) calculated from these land use designations.

\smallskip

\begin{figure}[ht]
\renewcommand\thefigure{S3}
\centering
\includegraphics[width=125mm]{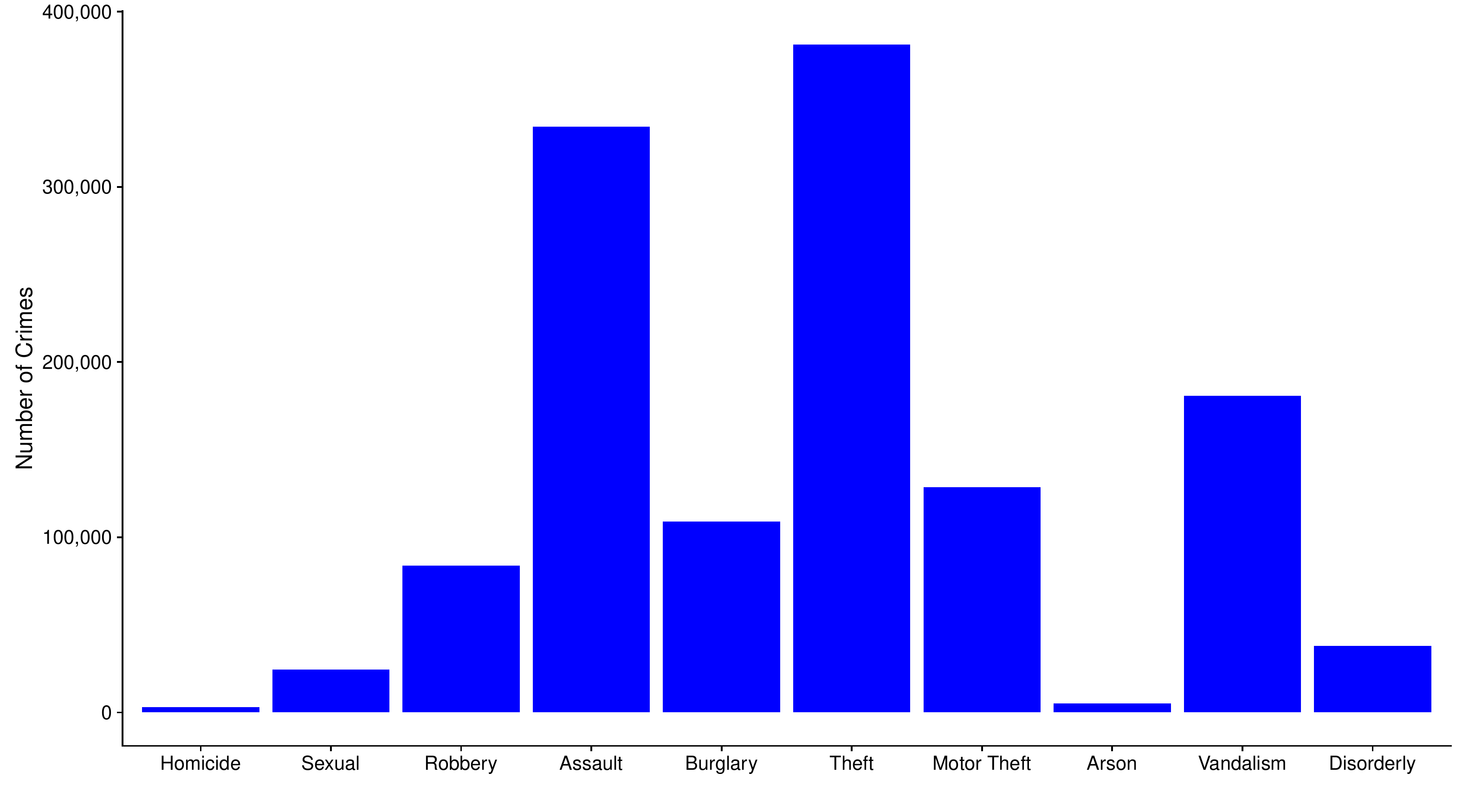}
\includegraphics[width=5.5in]{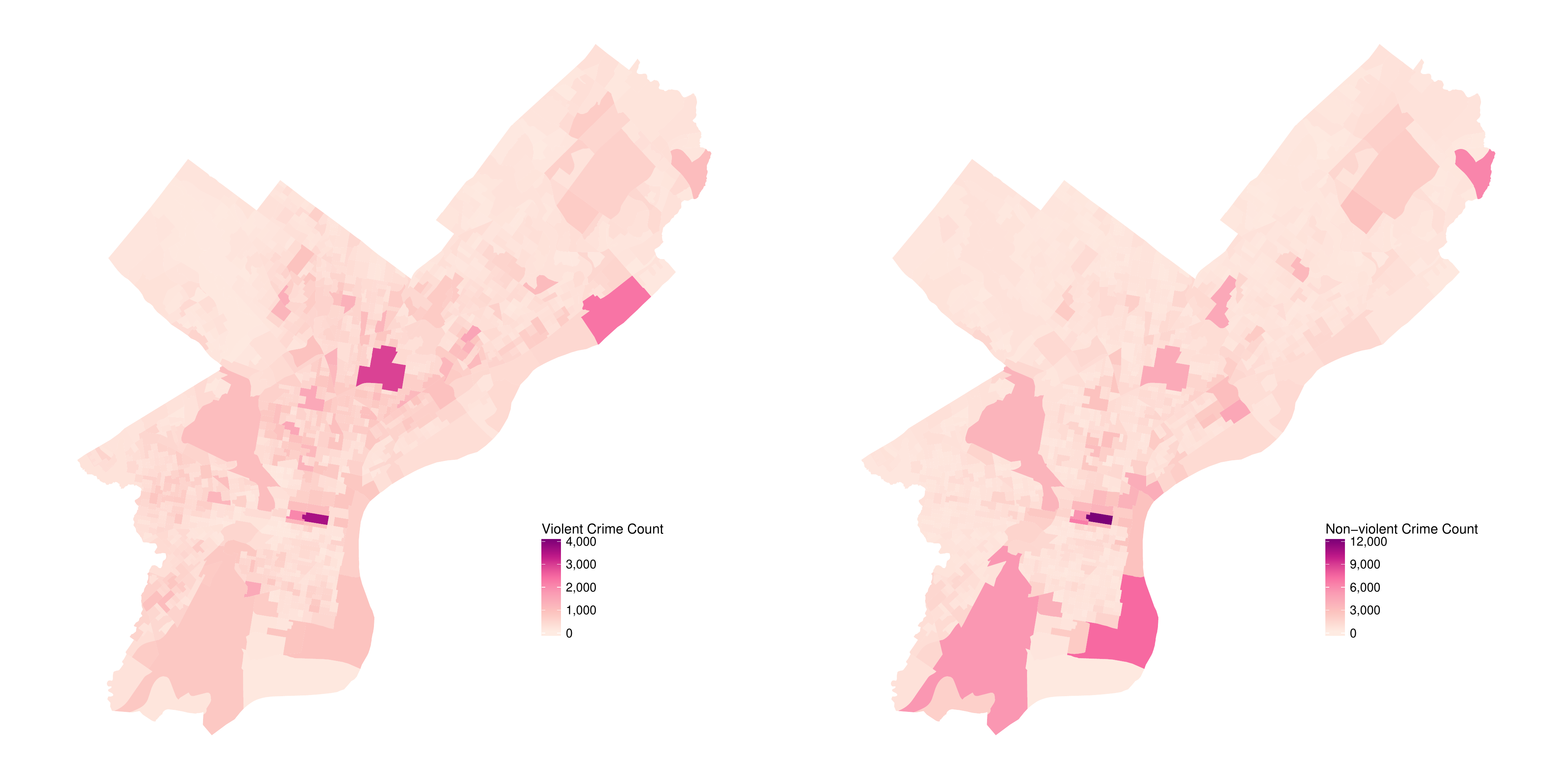}
\caption{{\bf Top:} Relative frequency of different crime types reported in Philadelphia from January 1, 2006 to December 31, 2015.   Number of Violent ({\bf Bottom Left}) and Non-Violent ({\bf Bottom Right}) Crimes by block group in Philadelphia over that same time period}
\label{map-crime}
\end{figure}

\noindent
Figure~\ref{map-crime} gives the relative frequency of each type of crime in our data as well as the spatial distribution by block group of violent vs. all crimes committed in Philadelphia from 2006-2015.   Note that these crime categories are roughly ordered in terms of severity, and that high severity crimes are much less frequent.   We see substantial heterogeneity in crime counts across the city with a large outlier count of both violent and non-violent crimes in the Market East block group of center city.   

\section{Exploring Neighborhood Factors Associated with Safety in Philadelphia} \label{section-eda}

\begin{figure}[ht!]
\renewcommand\thefigure{S4}
\centering
\includegraphics[width=5.5in]{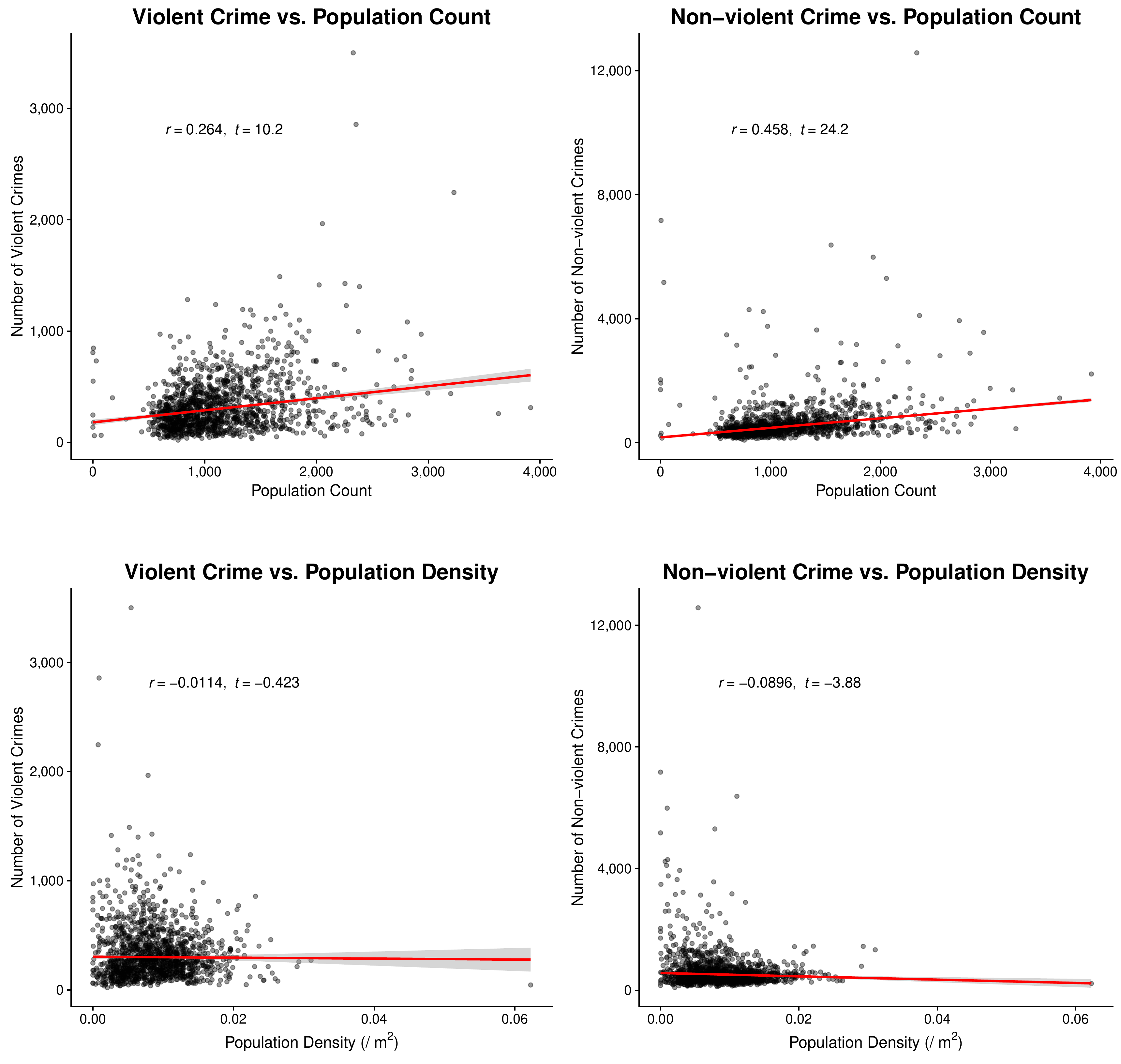}
\caption{Association between Safety and Population. Predictor variables are either population count (top row) or population density (bottom row).  Outcome variables are either violent crime counts (left column) or non-violent crime counts (right column).  Red lines (and grey bands) are the least squares lines (and confidence bands) from a robust regression that downweights outlying values. }
\label{panel-crime-vs-population}
\end{figure}

Figure~\ref{panel-crime-vs-population} plots the relationship between crime (either violent or non-violent) and population (either population count or population density).   We also provide the correlation and test statistic for the slope from a robust regression that downweights outlying values.

\smallskip

\begin{figure}[ht!]
\renewcommand\thefigure{S5}
\centering
\includegraphics[width=5.5in]{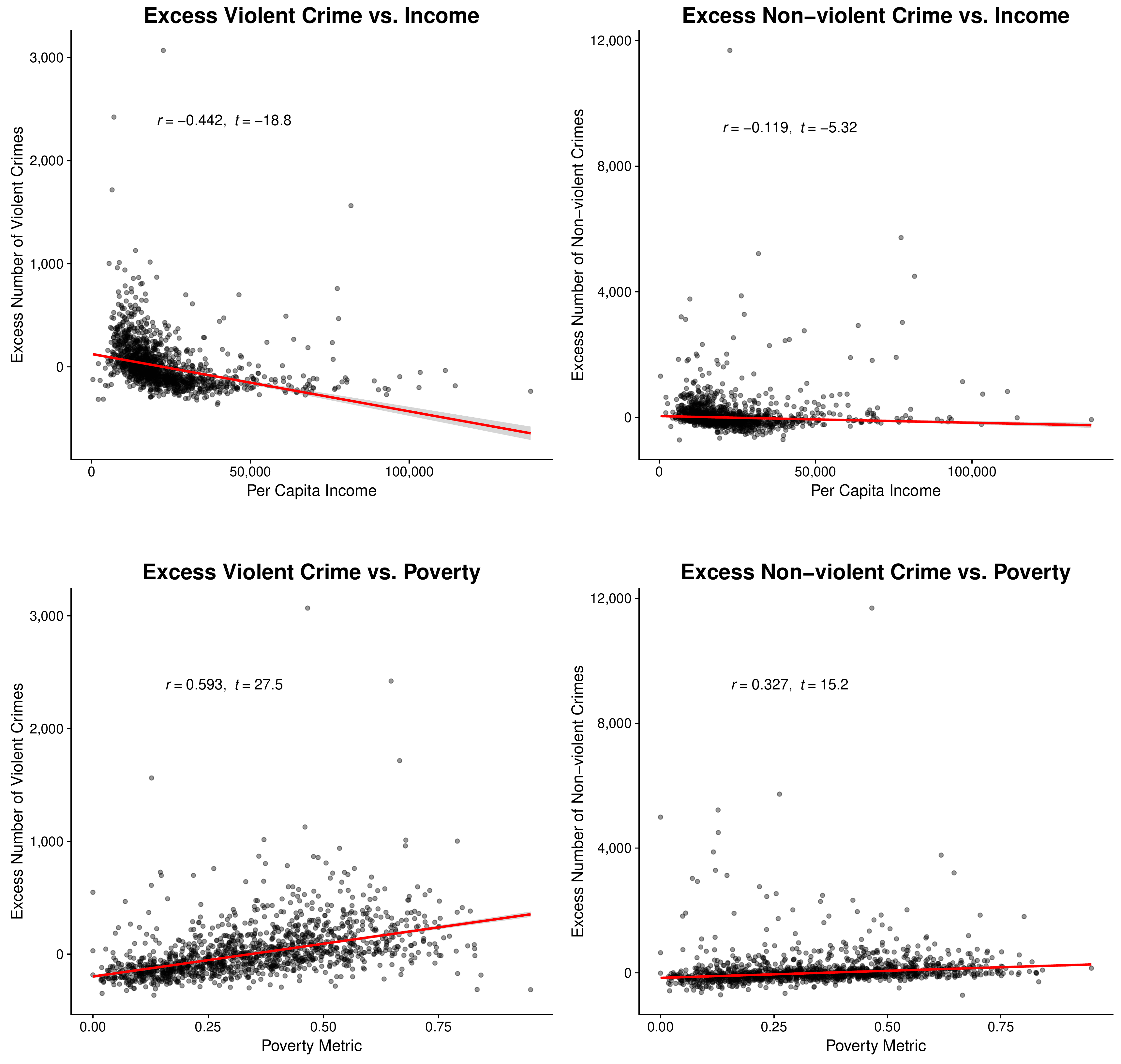}
\caption{Association between safety and economic measures. Predictor variables are either per-capita income (top row) or our poverty metric (bottom row). Outcome variables are either excess violent crime counts (left column) or excess non-violent crime counts (right column). Red lines (and grey bands) are the least squares lines (and confidence bands) from a robust regression that downweights outlying values.}
\label{panel-crime-vs-economic}
\end{figure}

\noindent
Figure~\ref{panel-crime-vs-economic} plots the relationship between excess crime (either violent or non-violent) and our economic measures 
(either per-capita income or poverty).  We also provide the correlation and test statistic for the slope from a robust regression that downweights outlying values.

\smallskip

\begin{figure}[ht!]
\renewcommand\thefigure{S6}
\centering
\includegraphics[width=5in]{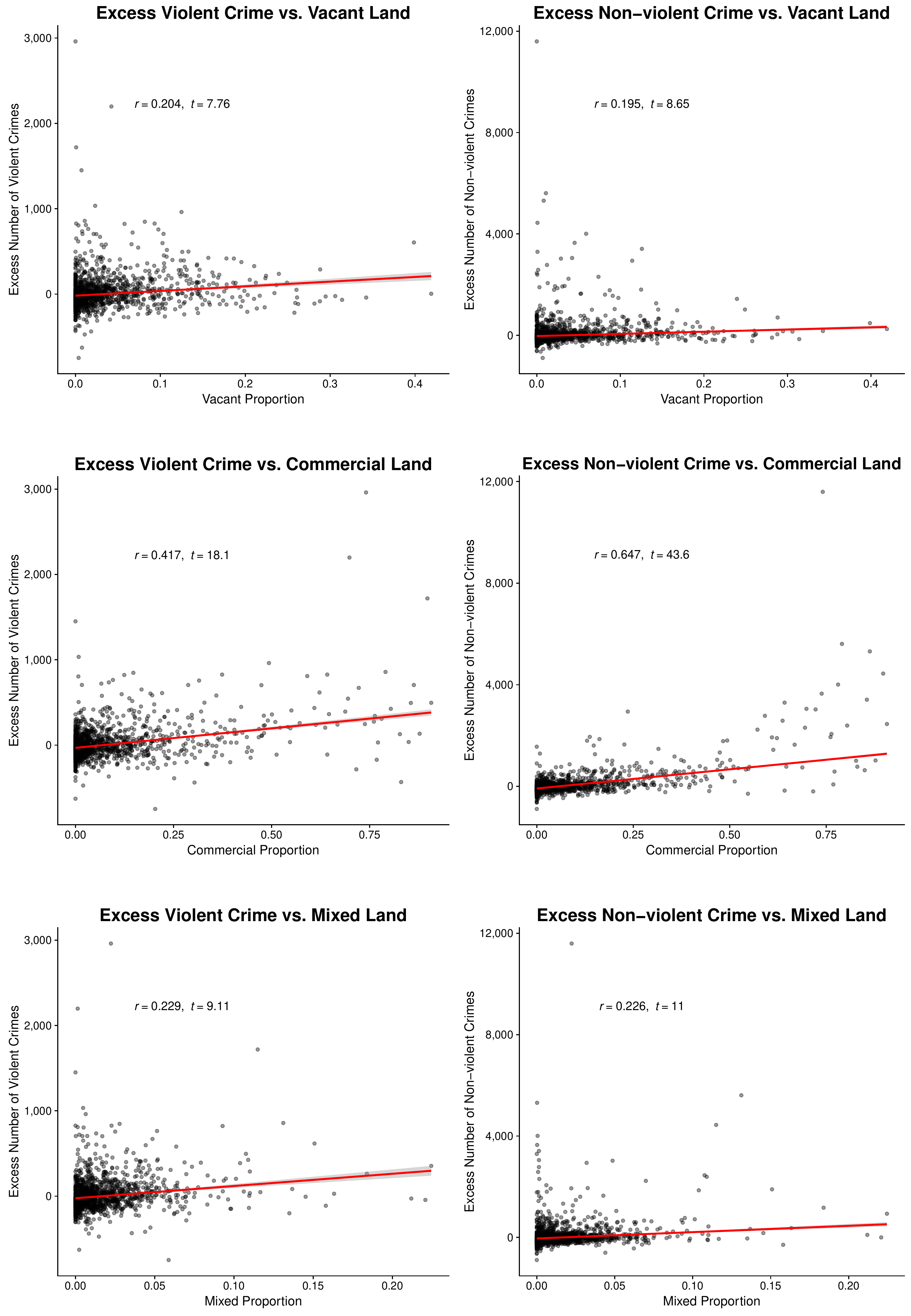}
\caption{Association between safety and land use vibrancy measures. Red lines (and grey bands) are the least squares lines (and confidence bands) from a robust regression that downweights outlying values.}
\label{panel-crime-vs-landuse}
\end{figure}

\noindent
Figure~\ref{panel-crime-vs-landuse} plots the relationship between excess crime (either violent or non-violent) and our land use measures (vacant proportion, commercial proportion and mixed use proportion).  We also provide the correlation and test statistic for the slope from a robust regression that downweights outlying values.

\section{Comparing ``Open Shorter" vs. ``Open Longer" Locations} \label{businessmatching}

The goal of our analysis in Section 5 of our paper is evaluating the association between measures of business vibrancy and safety at the local neighborhood level, while controlling for the characteristics of those neighborhoods by comparing pairs of locations within each block or within each block group.   In our main paper, we focus on comparisons of `high crime" vs. ``low crime" locations.  In this supplemental section, we examine block groups with one location that has ``open longer" businesses and another location that has ``open shorter" businesses.   We then examine these within-block-group pairs to see if there are differences in crime between ``open shorter" vs. ``open longer" locations. 

Specifically, for each of our ten business types, we identify block groups that contain a pair of businesses (of that type) where one of those businesses has long opening hours and the other business has short opening hours.  We define a business as having long opening hours if its total opening hours are above the $75$th percentile for businesses of that type.  Similarly, we define a business as having short opening hours if its total opening hours are below the $25$th percentile for businesses of that type. We further restrict ourselves to block groups where the pair of businesses are at least 140 meters apart, which is roughly the size of a Philadelphia city block.  Only a small subset of the block groups in Philadelphia will contain such a valid pair of locations. 

For each block group with a valid pair of locations, we then count the number of crimes that occurred within a 70 meter radius around both the long opening hour business location and then we calculate a matched pairs mean differences in crime around the short opening hour business minus long opening hour businesses in each within-block group pair.  If businesses that are active (open for a longer period) help to deter crime and promote safety, then these differences in crime should be positive.   

\begin{figure}[ht!]
\renewcommand\thefigure{S7}
\centering
\includegraphics[width=5.5in]{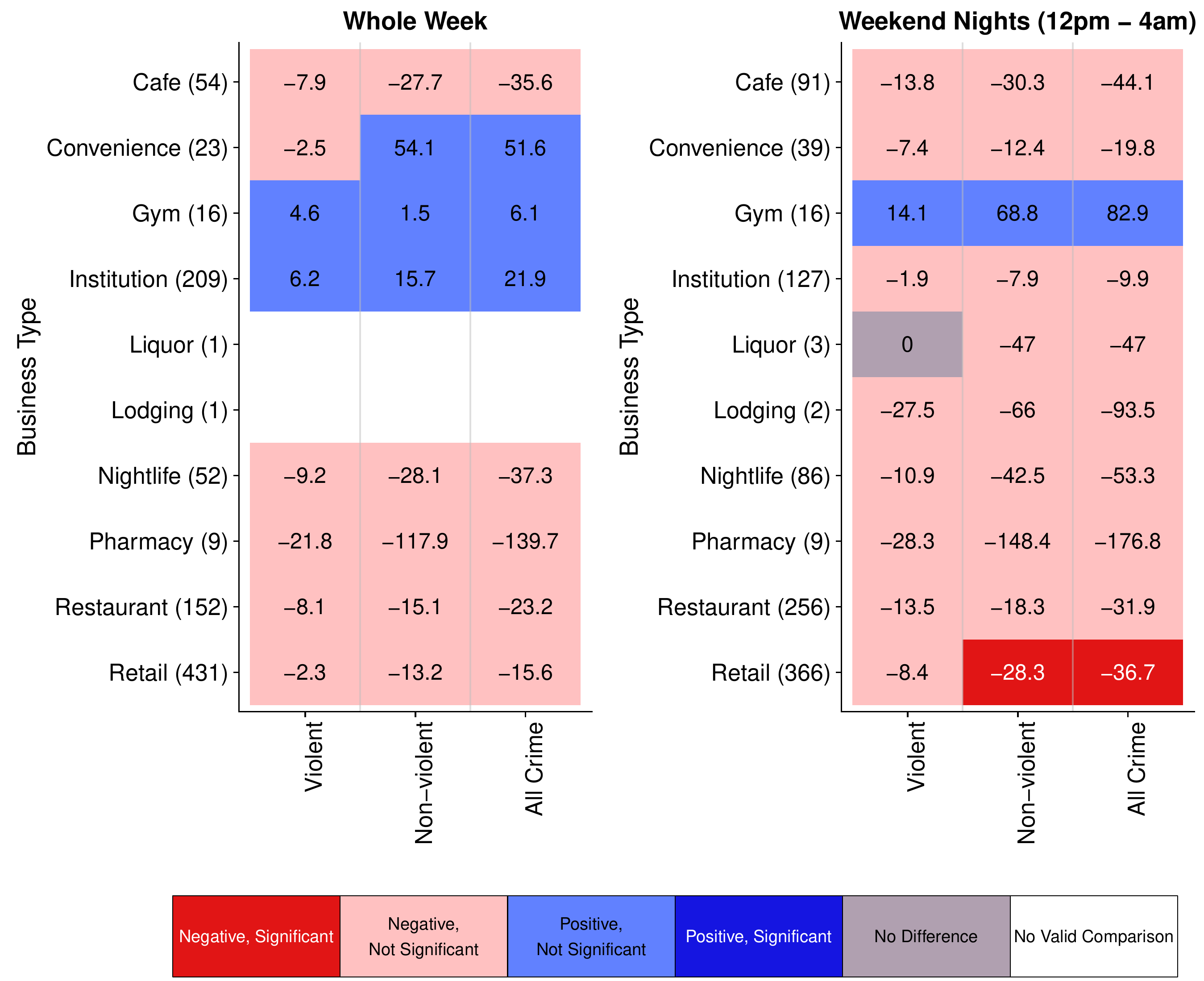}
\caption{Matched pair mean differences in crime between short opening and long opening hour businesses, calculated separately for each combination of crime type and business type.  Different panels are used to display the mean differences calculated over the entire week vs. just weekday evenings vs. just weekend nights. The significance threshold of $p = 0.05$ was Bonferroni-adjusted to account for multiple comparisons. Values in parentheses are the number of block groups with valid comparisons for that business type}
\label{fig.short.vs.long}
\end{figure}

In Figure~\ref{fig.short.vs.long}, we display the matched pair mean differences in crime between short opening and long opening hour businesses of each business type separately.   We calculate different matched pair mean differences for only violent crimes, only non-violent crimes and all crimes.  The significance threshold for these t-statistics was Bonferroni-adjusted to account for the number of comparisons being tested.  We also examine these comparisons for the entire week versus just the high crime {\it weekend nights} time window.  

In Figure~\ref{fig.short.vs.long}, we see mostly negative differences (red) which imply that more crimes are occurring around the business location with longer open hours, especially nightlife locations and restaurants.  A notable exception are gyms, which show positive differences that imply fewer crimes occurring around the gym with longer open hours.   Recall that our preliminary hypothesis was that greater business vibrancy would be associated with fewer crimes around those vibrant locations relative to less vibrant locations in the same block group.    The results in Figure~\ref{fig.short.vs.long} for gyms does show a trend in this expected direction, but the results for several other business types goes against that hypothesis.  That said, there are not many differences in Figure~\ref{fig.short.vs.long} that are statistically significant.  To a large extent, the lack of significance is driven by the small sample sizes in these comparisons.

\end{document}